 \documentclass[final,3p,times]{elsarticle}
 \usepackage{graphics}
\usepackage{amssymb}
\usepackage{amsmath}
\usepackage{cases}
 \biboptions{comma,square,compress,numbers,sort}

\journal{Physics Letters B}

\begin{document}

\begin{frontmatter}



\title{Localization of $q-$form fields on $AdS_{p+1}$ branes}


\author{Chun-E Fu$^1$}
\author{Yu-Xiao Liu$^2$\corref{cor1}}
  \ead{liuyx@lzu.edu.cn}
  \cortext[cor1]{The corresponding author.}
\author{Heng Guo$^3$}
\author{Feng-Wei Chen$^2$}
\author{Sheng-Li Zhang$^1$}

\address{$^1$School of Science, Xi'an Jiaotong University, Xi'an 710049, P. R. China\\
$^2$Institute of Theoretical Physics, Lanzhou University, Lanzhou 730000, P. R. China\\
$^3$School of Science, Xidian University, Xi'an 710071, P. R. China}

\begin{abstract}

In this paper, we investigate localization of a free massless $q-$form bulk field on thin and thick $AdS_{p+1}$ branes with codimension one. It is found that the zero mode of the $q-$form field with $q>(p+2)/2$ can be localized on the thin negative tension brane, which is different from the flat brane case given in [JHEP 10 (2012) 060]. For the thick $AdS_{p+1}$ branes, the $q-$form field with $q>(p+2)/2$ also has a localized zero mode under some conditions. Furthermore, we find that there are massive bound KK modes of the $q-$form field, which are localized on this type $p-$branes.
\end{abstract}


\end{frontmatter}

% \linenumbers

\section{Introduction}

Extra dimension and brane-world theories have received more and more attention since the last century,  especially when the Arkani-Hamed-Dimopoulos-Dvali (ADD) \cite{Antoniadis1998} and  Randall-Sundrum (RS) \cite{Randall1999a,Randall1999b} brane-world models were brought up, as they opened a new avenue to solve the long-standing hierarchy problem and the cosmology problem \cite{Rubakov2,ArkaniHamed1998rs,Cosmological2000,PhysRevLett.86.4223,coscon2009,Neupane2010ey,
BraneFR2012,ThickBrane20093}. Then many brane-world models were set up \cite{ThickBraneDewolfe,Gregory:2000jc,Kaloper2000jb,PhysRevD.66.024024,ThickBrane2002,ThickBrane2003,Blochbrane2004,
ThickBraneBazeia2006,ThickBrane20071,ThickBrane20082,ThickBrane20093,
KeYang2009Weyl,ThickBraneZhongYuan2011JHEP,Liu:2012rc,
Liu:2012mia,BraneFR2012,Bazeia:2013uva,Gherghetta:2000jf,Zhong:2010ae}.

In this paper, we will consider an interesting brane-world model, the $AdS_{p+1}$ branes with codimension one, which have $p$ spacial dimensions and an effective non-zero cosmological constant. These branes could have some realistic applications if the branes have $3$ infinite large dimensions (which are those we can feel) and $p-3$ finite small enough dimensions with topology $S^1 \times S^1 \times ... \times S^1 = T^{p-3}$ (thus can not be seen. According to string/M theory, there are 6 or 7 extra dimensions).
%And the non-zero cosmological constant on the branes is more consistent with
%the astronomical observation nowadays than the flat brane.
It is known that the $AdS$ space has many unusual properties \cite{Witten:1998qj}, and plays a center role in the $AdS/CFT$ correspondence \cite{Gubser:1998bc,Maldacena:1997re}.

In the brane-world theory, investigating the KK modes of various fields is an important and interesting work \cite{Grossman:1999ra,Bajc:1999mh,Gremm:1999pj,Chang1999nh,RandjbarDaemi:2000cr,Kehagias:2000au,
Oda2001,Ringeval:2001cq,Ichinose:2002kg,Koley:2004at,Davies:2007xr,
Liu2008WeylPT,Zhao:2010mk,Jones:2013ofa,Xie2013rka,Cembranos:2013qja,Costa:2013eua,Duff2000se}. Because the information of extra dimensions is encoded in these KK modes. And the most important thing is that the massless modes (zero modes) are in fact the fields which have existed on the brane. Thus, the zero modes of ordinary fields (gravity, scalar, vector, and fermion et al.) localized on the brane can rebuild the effective gravity and standard model in our world \cite{Chang1999nh}.

It is known that, in the string/M theory, the $q-$form fields are related to closed strings, which play an important role in the $D-$ or $M-$brane \cite{Ma:2012hc,Ho:2013paa}. As the brane-world theory is motivated by the string/M theory, we naturally ask what about the $q-$form fields on the $AdS_{p+1}$ branes. On the other hand, although the $q-$form fields with $q>1$ are dual to the scalar field in four-dimensional spacetime, they denote  new types of particles in higher-dimensional spacetime. Thus,  we investigate in this paper the KK modes of the $q-$form fields on the $AdS_{p+1}$ branes.

In fact,  the $q-$form fields have been investigated in some literatures. For example, in Refs. \cite{PRLKR2002,Mukhopadhyaya2004,QformRS,qRSdilation-eprintv2,QformThickRS}, the KK modes of the $2-$form and $3-$form fields were discussed in the RS brane model, and it was found that the zero modes of these fields cannot be localized on the brane, unless they are coupled with a dilaton field. In Ref. \cite{Fu2012sa}, some authors of the current work considered any form ($q-$form) field on the flat $p-$branes with $p$ the spatial dimensions of the brane, and found that only the lower form fields (with $q<(p-1)/2$ or $q<p/2$) can be localized on the flat $p-$branes.

In this work we will find that the higher form fields (with $q>(p+2)/2$) can be localized on the thin negative tension $AdS_{p+1}$ brane and the thick $AdS_{p+1}$ brane, which is an interesting result. For example, on the thin $AdS_4$ brane with negative tension and the thick one, there is a $3-$form field that can be localized.
This may be useful for some unknown problems such as the cosmological constant problem or dark energy problem \cite{PhysRevD.88.123512}.
In fact, in Ref. \cite{PhysRevLett.86.4223}, the authors had considered a 3-form field with a nonstandard action to solve the cosmological constant problem in the five-dimensional RS model.

This paper is organized as follows. We first simply review the localization mechanism for the KK modes of the $q-$form fields in Sec. \ref{localizationmechanism}. Then we try to find the solutions of  $AdS_{p+1}$ branes in Sec. \ref{warpedpbrane}, and investigate the localization of the KK modes on these branes. Finally we give a brief conclusion.

\section{$q-$form fields and their KK modes}\label{localizationmechanism}

In Ref. \cite{Fu2012sa}, some authors of the current paper proposed a localization mechanism for {a free massless $q-$form field}, which can {be applied to} any RS-like braneworld model. Here we briefly review the localization mechanism.

{The action for a free massless $q-$form field $X_{M_1M_2\cdots M_{q}}$, which is completely antisymmetric, in a curved $D-$dimensional spacetime is}
\begin{eqnarray}\label{q-action}
S_q=\int d^D x\sqrt{-g}\;Y_{M_1M_2\cdots  M_{q+1}}Y^{M_1M_2\cdots  M_{q+1}}, \label{action_S_q}
\end{eqnarray}
where $Y_{M_1M_2\cdots M_{q+1}}=\partial_{[M_1}X_{M_2\cdots M_{q+1}]}$ is the field strength.
The $q-$form field may have some localized KK modes on lower dimensional branes (such as the $p-$brane with $D=p+2$) embedded in the $D-$dimensional spacetime with the metric $ds^2=\text{e}^{2A(w)}(\hat{g}_{\mu\nu}(x^\lambda)dx^{\mu} dx^{\nu}+dw^2)$. This can be found by the KK decomposition and KK reduction.

There exists a gauge freedom $\delta X_{M_1\cdots M_{q}}=\partial_{[M_1}\Lambda_{M_2\cdots M_{q}]}$ for the $q-$form field. So we can choose a simple gauge $X_{\mu_1\cdots \mu_{q-1}w}=0$. Decomposing the remained components of the $q-$form field as \begin{equation}
  X_{\mu_1\mu_2...\mu_q}(x^\mu,w)=\sum_n
\hat{X}^{(n)}_{\mu_1\mu_2\cdots \mu_q}(x^\mu)U^{(n)}(w) \text{e}^{(2q-p)A/2},
\end{equation}
where $\hat{X}^{(n)}_{\mu_1\mu_2\cdots \mu_q}(x^\mu)$ is only relative to the {brane coordinates} and $U^{(n)}(w)$ {is the function of} the extra dimension, {and substituting} the decomposition into the equation of motion for the $q-$form field, {one will find that the KK mode}  $U^{(n)}(w)$ satisfies the following Schr\"{o}dinger-like equation:
 \begin{eqnarray}
\big[ -\partial^2_w+ V_q(w)\big]U^{(n)}(w)
  =m_n^2 U^{(n)}(w),
  \label{SchEq}
\end{eqnarray}
where $m_n$ {is the mass of the KK mode}, and the effective potential {$V_q(w)$} takes the form
\begin{eqnarray}\label{V}
V_q(w)=\frac{(p-2q)^2}{4} (\partial_w A(w))^2 +\frac{p-2q}{2}\partial^2_w A(w).
\label{Veff}
\end{eqnarray}

For the above Schr\"{o}dinger-like equation {(\ref{SchEq})}, we need an orthonormality condition to seek the exact solution for $U^{(n)}$ and $m_n$. As our aim is to find the KK modes which can be localized on the brane, the orthonormality condition can be obtained by {the} KK reduction for the action (\ref{q-action}):
\begin{eqnarray}
S &=& {\sum_{n,m}} \int d^{p+1}x \;\sqrt{-\hat{g}}
\bigg(\hat{Y}^{(n)}_{\mu_1\mu_2...\mu_{q+1}}(x^\mu)\hat{Y}^{(m)\mu_1\mu_2...\mu_{q+1}}(x^\mu)\nonumber\\
&&+\frac{1}{q+1}m_n^2\hat{X}^{(n)}_{\mu_1\mu_2...\mu_q}(x^\mu)
\hat{X}^{{(m)}\mu_1\mu_2...\mu_q}(x^\mu)\bigg)
\int dw\; U^{(m)}(w)U^{(n)}(w).
\end{eqnarray}
If {$U^{(m)}(w)$ satisfy the following orthonormality condition:}
\begin{equation}\label{NormCondi}
  \int U^{(m)}(w)U^{(n)}(w)dw=\delta_{mn},
\end{equation}
{we will get the effective action for a series of $q-$form fields on the brane}:
\begin{equation}
  S_{eff}= \sum_n \int d^{p+1}x\sqrt{-\hat{g}}
\bigg(\hat{Y}^{(n)}_{\mu_1\mu_2...\mu_{q+1}}\hat{Y}^{(n)\mu_1\mu_2...\mu_{q+1}}
+\frac{m_n^2}{q+1}\hat{X}^{(n)}_{\mu_1\mu_2...\mu_q}\hat{X}^{{(n)}\mu_1\mu_2...\mu_q}
\bigg).
\end{equation}
This effective action is just for the KK modes localized on the brane, and the localization condition is the orthonormality condition (\ref{NormCondi}).

Thus with the Schr\"{o}dinger-like equation (\ref{SchEq}) and the orthonormality condition
(\ref{NormCondi}), we can investigate the KK modes of {a $q-$form field} on branes. There are usually two kinds of KK modes localized on the brane, i.e., the massless (zero mode) and the massive bound {KK modes}. The zero mode is actually irrelative to the extra dimension {(from $U_0\propto\text{e}^{(p-2q)A/2}$, we have $X^{(0)}_{\mu_1\mu_2...\mu_q}(x^\mu,w)= \hat{X}^{(0)}_{\mu_1\mu_2...\mu_q}(x^\mu)$)}, and so it can be regarded as the matter that has existed on the brane. While the massive bound modes are relative to the extra dimension, and the mass spectrum carries the information of the extra dimension. Thus, we should find the mass spectrum of the KK modes. This requires to discuss the effective potential $V_q$ (\ref{Veff}), which is
decided by the geometry of the background spacetime. In this paper, we will consider $AdS_{p+1}$ branes and investigate the KK modes of $q-$form fields on them.

We will first find solutions of the thin and thick $AdS_{p+1}$ branes with codimension one. For the thin brane, we can get analytic solutions. For the thick ones, the solutions are depended on the special background scalar potential $V(\phi)$.  In this paper, we will consider the simple but typical $\phi^4$ potential, and use the numerical method to obtain the solutions with the initial and boundary conditions.
However, we will not give the particular results in this paper, because the initial and boundary conditions are enough for us to analyze the behavior of the effective potentials for the KK modes.

\section{$AdS_{p+1}$ branes}\label{warpedpbrane}

We assume that the line element for the $D=p+2$ dimensional spacetime is
\begin{eqnarray}\label{metric1}
  ds^2=\text{e}^{2A(r)}\hat{g}_{\mu\nu}dx^{\mu} dx^{\nu}+dr^2,
\end{eqnarray}
where $r$ denotes the physical extra dimension perpendicular to the $p-$brane, and $\hat{g}_{\mu\nu}$ is the induced metric on the brane. In this work, we will focus on the $AdS_{p+1}$ branes, for which the metric $\hat{g}_{\mu\nu}$ is given by
\begin{eqnarray}
  \hat{g}_{\mu\nu}dx^{\mu} dx^{\nu}&=&\text{e}^{2Hx_p}(-dt^2+dx_1^2+\dots+dx_{p-1}^2)+dx_p^2.
\end{eqnarray}
Here, $H$ is a parameter associated with the effective cosmology constant $\Lambda_p$ on the $p-$brane by $\Lambda_p=- p\;H^2$ \cite{Mannheim2005}.
Then the nonvanishing components of the Einstein tensor are turned out to be
\begin{eqnarray}\label{adsG}
  G_{\mu\nu}&=& \frac{1}{2}p
  %\;\text{e}^{2H x_p}
      \Big[2A'' +(p+1)A'^2 +(p-1)\text{e}^{-2A}H^2
      \Big] \hat{g}_{\mu\nu}.
\end{eqnarray}
In this paper, the prime denotes the derivative with respect to the physical extra dimension coordinate $r$.

Then with the matter field we can get the Einstein equations. In the following, we will consider the thin and thick branes, in which the {``matters" are} made up by the brane tension $\sigma$ (the energy density per unit area) and a scalar field $\phi$, respectively.

\subsection{Thin $AdS_{p+1}$ branes}

We first consider the thin brane case, for which the action is
\begin{eqnarray}
  S=\frac{1}{2}\int d^D x\sqrt{-g}(R{-p\;\Lambda})+\int d^{D-1}x\sqrt{-g^{(b)}}(-\sigma), \label{actionthinbrane}
\end{eqnarray}
where {$\Lambda$} is a negative cosmology constant and $g^{(b)}_{\mu\nu}$ is the induced metric on the brane. From the action (\ref{actionthinbrane}) we obtain the Einstein equations:
\begin{eqnarray}
  {G_{MN}+\frac{1}{2}p\Lambda g_{MN}}
  =-\frac{1}{\sqrt{g_{rr}}}\delta(r)\sigma\delta^\mu_M\delta^\nu_N g_{\mu\nu},
\end{eqnarray}
i.e.,
\begin{eqnarray}
  G_{00}
  &=&-\Big[\frac{1}{\sqrt{g_{rr}}}\sigma\delta(r){+\frac{1}{2}p\Lambda}\Big] g_{00}, \nonumber\\
  G_{rr}
  &=&{-\frac{1}{2}p\Lambda g_{rr}}.
\end{eqnarray}

Considering the expressions of the Einstein tensor for the $AdS_{p+1}$ brane,
the above Einstein equations are rewritten as:
\begin{eqnarray}
A'^2&=&b^2- H^2\text{e}^{-2A}\label{eq1},\\
A''&=&-\frac{\sigma}{p}\delta(r)+ H^2\text{e}^{-2A},\label{eq2}
\end{eqnarray}
where $b={\sqrt{-\Lambda/(p+1)}}$.
From the above equations, we can obtain the solution of the $AdS_{p+1}$ brane, which will be shown below.

Our aim is to investigate the KK modes of the $q-$form field; {to this end we need} to discuss the effective potential $V_q$ for the KK modes. With the coordinate transformation $dw=\text{e}^{-A}dr$, the  potential $V_q$  in Eq. (\ref{V}) can be expressed as the function of $r$:
\begin{eqnarray}
  V_q(w(r)) &=& \frac{p - 2q}{4}\text{e}^{2A(r)}\Big[(p-2q+2) A'(r)^2
                                   + 2A''(r)\Big].
\end{eqnarray}
It is clear that with the Einstein equations (\ref{eq1}), the behavior of the effective potential $V_q(w(r))$ at $r=0$ and $r\rightarrow r_{\text{bou}}$ is
\begin{eqnarray}\label{Vthin0}
  V_q(r\rightarrow 0) &=& -\frac{(p-2q)\sigma}{2p}\delta(0),\label{Vqthin0}\\
  V_q(r\rightarrow r_{\text{bou}})&=&  \frac{p-2q}{4} \Big[(p-2q+2)b^2
  \;\text{e}^{2A(r\rightarrow r_{\text{bou}})}-(p-2q)H^2 \Big]\label{Vthininf},
\end{eqnarray}
where $r_{\text{bou}}$ is the boundary of the extra dimension. If $V_q(r\rightarrow 0) <0$ and $V_q(r\rightarrow r_{\text{bou}})>0$, there will exist a zero KK mode $U_0 \propto \text{e}^{(p-2q)A/2}$. We then can check whether it can be localized on the brane by the orthonormality condition (\ref{NormCondi}):
\begin{equation}\label{zerocondition}
  \int U_0^2\;dw=\int U_0^2\;\text{e}^{-A}\;dr{\propto}\int \text{e}^{(p-2q-1)A}\;dr\rightarrow cons.
\end{equation}

In order to check the above condition, we will look for the solution of the warp factor by solving the Einstein equations. And from (\ref{Vqthin0}), we see that for the $AdS_{p+1}$ branes with negative tension $\sigma$, $V_q(r\rightarrow 0) <0$ can be satisfied for $q>p/2$, which means that there may exist the $q-$form fields with $q>p/2$ on the branes, and this is different with that on the flat $p-$branes \cite{Fu2012sa}. While for branes with positive tension, the $q-$form fields on the branes are the ones with $q<p/2$, which is similar to that on the flat $p-$branes. So in the following we only focus on the thin $AdS_{p+1}$ with $\sigma<0$, and discuss the existence and localization conditions for the zero mode of the $q-$form fields in detail.

For the $AdS_{p+1}$ brane with negative tension $p-$brane, the solution is
\begin{eqnarray}\label{thinadswarp1}
  A(r)&=&\ln{\Big(\frac{H}{b}\cosh{(b\;|r| + c)}\Big)},\\
  \sigma&=& -2p\sqrt{b^2-H^2}.
\end{eqnarray}
Here $c=\text{arccosh}(b/H)$, and the brane cosmology constant $\Lambda^{{AdS_{p+1}}}_{\text{brane}}$ is given by
\begin{eqnarray}
 \Lambda^{{AdS_{p+1}}}_{\text{brane}}=-p\;H^2=p
                 \bigg(\frac{\Lambda}{p+1} + \frac{\sigma^2}{4 p^2}\bigg),
\end{eqnarray}
where the bulk cosmology constant should satisfy the following condition:
\begin{equation}\label{backlambda1}
\Lambda <- \frac{(p+1)}{4 p^2}{\sigma^2}, ~~~\text{or}~~~
\Lambda <- (p+1)H^2.
\end{equation}
From the warp factor, it can be seen that the physical extra dimension for $AdS_{p+1}$ is infinite, i.e., $r_{\text{bou}}\rightarrow \infty$, and the behavior of the warp factors at infinity is $A(r\rightarrow\infty)\rightarrow\text{e}^{b\;r}$.

According to Eqs. (\ref{Vthin0}) and (\ref{Vthininf}), the values of the effective potential at $r\rightarrow0$ and at infinity are obtained:
\begin{eqnarray}
  V_q(r\rightarrow0)&=&-\frac{(p-2q)\sigma}{2p}\delta(0),\\
  V_q(r\rightarrow \infty)&\rightarrow&  {\frac{b^2}{4}(p-2q)(p-2q+2)\;\text{e}^{2b\;r}},
\end{eqnarray}
which determines whether a localized zero mode exists.
If there is a zero mode, we can check whether it can be localized on the brane:
\begin{equation}
  \int_{-\infty}^{\infty} U_0^2\;dw=\int_{-\infty}^{\infty} \text{e}^{(p-2q-1)A}dr\rightarrow { 2\int_0^{\infty}\text{e}^{b(p-2q-1)r}dr < \infty}.
\end{equation}
Then it is clear that only for
\begin{equation}\label{zero1}
  q>\frac{p-1}{2}
\end{equation}
the above integral is finite, and the zero mode can be localized on the brane. With the localization condition (\ref{zero1}), we seek for the existence condition for the zero mode, i.e., $V_q(r\rightarrow0)<0$ and $V_q(r\rightarrow \infty)>0$, which requires that
\begin{eqnarray}
  q>\frac{p}{2}+1,
\end{eqnarray}
Remember that only the $q-$form fields with $q<p+1$ play a role in the $p-$brane. Thus finally the conditions for the zero mode to be existence and localized are:
\begin{eqnarray}\label{zero2}
  \frac{p}{2}+1<q<p+1,
\end{eqnarray}
Thus for the thin $AdS_{p+1}$ brane with negative tension, there exists a localized zero mode with the condition (\ref{zero2}). For example, on the thin $AdS_4$ brane, a $3-$form field can be localized.
This result is different with that in the flat thin $p-$brane \cite{Fu2012sa}.

\subsection{Thick $AdS_{p+1}$ branes}

In this subsection, we consider the thick $AdS_{p+1}$ branes generated by a scalar field with scalar potential, which are more realistic. { The action of such a} system is
\begin{eqnarray}
  S&=&\int d^Dx\;\sqrt{-g}\bigg(\frac{1}{2}R-\frac{1}{2}g^{MN}\partial_M\phi\partial_N\phi-V(\phi)\bigg),
\end{eqnarray}
where $\phi=\phi(w)$ is the scalar field, and $V(\phi)$ is a potential which generates the brane. We can obtain the Einstein { equations} using the metric { ansatz} (\ref{metric1}):
\begin{eqnarray}
  R_{MN}-\frac{1}{2}g_{MN}R=g_{MN}\Big(-\frac{1}{2}\phi'^2-V\Big)
  +\partial_M\phi\partial_N\phi,
\end{eqnarray}
which read as
\begin{eqnarray}
  G_{00}&=&g_{00}(-\frac{1}{2}\phi'^2-V),\\
  G_{rr}&=&\frac{1}{2}\phi'^2-V.
\end{eqnarray}
With the expressions of { the Einstein tensor} (\ref{adsG}),  we { get}
\begin{eqnarray}
   A'^2&=&-\text{e}^{-2A}H^2+\frac{1}{p(p+1)}\big(\phi'^2-2V\big),\label{thickeq1}\\
   A''&=&-p\;\text{e}^{-2A}H^2-(p+1)A'^2-2V/p\label{thickeq2}.
\end{eqnarray}
The field equation for the scalar field is
\begin{eqnarray}
  \frac{dV}{d\phi}=(p+1)A'\phi'+\phi''\label{eq3}.
\end{eqnarray}

In order to find the solutions of these equations (\ref{thickeq1})-(\ref{eq3}), we use the numerical method. Because the three equations are not independent, we suppose the form of the background scalar potential is
\begin{equation}
  V(\phi)=v_0+g_1\phi^2+g_2\phi^4,
\end{equation}
where $v_0, g_1,$ { and $g_2$} are constants. As the equations of motion are { second order, we} need four initial or boundary conditions. We choose the conditions as
\begin{equation}\label{initialcondition}
  A(0)=0,~~~~A'(0)=0,~~~~\phi(0)=0,~~~~\phi'(r_{\text{bou}})=0.
\end{equation}
Here the $r_{\text{bou}}$ is the boundary of the extra dimension, which may be infinite or finite.

What should be noted from Eq. (\ref{thickeq1}) is that with the initial conditions $A(0)=0$ { and} $A'(0)=0$, there is a constraint between the parameters $v_0$, { $H$,} and the spacial dimension $p$:
\begin{eqnarray}
  H^2>-\frac{2v_0}{p(p+1)}\label{condthickads},
\end{eqnarray}
where the constant $v_0$ can be negative or positive for $AdS_{p+1}$ brane.

Also with the initial and boundary conditions, we can analyze the behavior of the effective potential { for the KK modes of the $q-$form field, $V_q(w(r))$,} at $r=0$ and at $r\rightarrow r_{\text{bou}}$ from the { equations} of motion { of the background spacetime (\ref{thickeq1})-(\ref{eq3})}. As the warp factor is $A(0)=0$, $A'(0)=0$, { and $A''(0)=(- p\;H^2-2v_0)/p$  at $r=0$},  { the value of the effective potential at $r=0$} is found to be
\begin{equation}\label{Vthick0}
  V_q(r\rightarrow0)=\frac{p-2q}{2}\xi,
\end{equation}
with $\xi=\Big(- p\;H^2-\frac{2v_0}{p}\Big)$. At $r\rightarrow r_{\text{bou}}$, as $\phi'(r\rightarrow r_{\text{bou}})=0$,  { it is easy to get}
\begin{equation}\label{Vthickinfity}
  V_q(r\rightarrow r_{\text{bou}})=\frac{(p-2q)}{4}\Big[ k{(p-2q+2)}\text{e}^{2A}- {(p-2q)}H^2\Big],
\end{equation}
where
\begin{equation}\label{k}
  k=-\frac{2V(\phi(r\rightarrow r_{\text{bou}}))}{p(p+1)}.
\end{equation}

From (\ref{Vthick0}), it can be seen that for $\xi>0$ and $q>p/2$, $V_q(r\rightarrow0)<0$, which suggests that there may be higher form fields with $q>p/2$ on the branes, and this is new result different with that on the thick flat $p-$branes \cite{Fu2012sa}. With the conditions (\ref{condthickads}), we find that for the $AdS_{p+1}$ brane, the value of $\xi$ can be positive with $v_0<0$.

For the $AdS_{p+1}$ case,  as $r\rightarrow r_{\text{bou}}$  {Eq.} (\ref{thickeq1}) can be reduced to $A'^2(\phi(r\rightarrow r_{\text{bou}}))=-\text{e}^{-2A(r_{\text{bou}})}H^2+k$, from which  {it is} seen that $k$ must be positive. The behavior of the warp factor at $r_{\text{bou}}$ is
\begin{eqnarray}
  A(r\rightarrow r_{\text{bou}})&=&\log{\big(\cosh{\sqrt{k}(r-c_1)}\big)}
   \rightarrow \sqrt{k}\;r.
\end{eqnarray}
We see that the { physical } extra dimension {is infinite}.

{For a zero mode of the $q-$form filed}, we can check whether the localization condition (\ref{zerocondition}) is satisfied. As the integral is
\begin{eqnarray}
  \int_{-\infty}^{\infty} U_0^2dw=  \int_{-\infty}^{\infty} \text{e}^{(p-2q-1)A}dr\rightarrow   2\int_0^{\infty}\text{e}^{\sqrt{k}(p-2q-1)r}dr,
\end{eqnarray}
it will be finite for
\begin{equation}\label{condtionthickadszero}
  q>\frac{p-1}{2}.
\end{equation}
Under the localization condition (\ref{condtionthickadszero}), we find the existence condition, which requires $V_q(r=0)<0$ and $V_q(r\rightarrow+\infty)>0$, i.e.,{
\begin{eqnarray}
  H^2&<&-\frac{2v_0}{p^2}~~\text{and}~~  q>\frac{p}{2}+1,
\end{eqnarray}
or
\begin{eqnarray}
  H^2&>&-\frac{2v_0}{p^2}~~\text{and}~~  q<\frac{p}{2}.
\end{eqnarray}}
{Considering the constrains (\ref{condthickads}) and (\ref{condtionthickadszero}), we finally get the following conclusion: under the condition for $H^2$: $-\frac{2v_0}{p(p+1)}<H^2<-\frac{2v_0}{p^2}$, the zero mode for the $q-$form field with $q>p/2+1$ can be localized on the $AdS_{p+1}$ brane; while under another condition: $H^2>-\frac{2v_0}{p^2}$, the zero mode for the $q-$form field with $(p-1)/2<q<p/2$ can also be localized on the brane. Because there is no integer solution of $q$ for $(p-1)/2<q<p/2$, we can not get a localized zero mode under the condition $H^2>-\frac{2v_0}{p^2}$.}

On the other hand, because the effective potential tends to infinite at $r\rightarrow \infty$, there are infinite massive bound KK modes.

\section{Conclusions}
\label{conclusion}

In this work, we investigated the KK modes of the $q-$form fields on the $AdS_{p+1}$ branes with codimension one, including in the thin and thick {$AdS_{p+1}$ ones. Through the KK decomposition and KK reduction, { we} found that the KK modes satisfy a Schr\"{o}dinger-like equation. By { analyzing} the equation under the orthonormality condition, we {finally obtained} the mass spectrum of the KK modes on these branes.

It was found that for the thin negative tension $AdS_{p+1}$ brane, there are localized zero modes on the brane for the $q-$form fields with $p/2+1<q<p+1$. For example, there exists a localized massless $3-$form field on the thin $AdS_4$ brane.

While for the thick $AdS_{p+1}$ branes, under some conditions between the parameters $H, v_0$ and $p$, the $q-$form fields with $p/2+1<q<p+1$ have localized zero modes. And there are massive bound KK modes, which are also localized on this type $p-$branes.

\chapter{\textbf{Acknowledgments}}

This work was supported by the National Natural Science Foundation of China (Grants No. 11375075 and No. 11305119),
and the Fundamental Research Funds for the Central Universities (Grants No. lzujbky-2013-18 and No. lzujbky-2013-227).
%\bibliographystyle{model1a-num-names}
%\bibliographystyle{elsarticle-num}
%\bibliography{D:/jabref/library/articles/articles}



%\bibliography{E:/Recently_work/References/references_new}

\begin{thebibliography}{59}
\expandafter\ifx\csname natexlab\endcsname\relax\def\natexlab#1{#1}\fi
\providecommand{\url}[1]{\texttt{#1}}
\providecommand{\href}[2]{#2}
\providecommand{\path}[1]{#1}
\providecommand{\DOIprefix}{doi:}
\providecommand{\ArXivprefix}{arXiv:}
\providecommand{\URLprefix}{URL: }
\providecommand{\Pubmedprefix}{pmid:}
\providecommand{\doi}[1]{\href{http://dx.doi.org/#1}{\path{#1}}}
\providecommand{\Pubmed}[1]{\href{pmid:#1}{\path{#1}}}
\providecommand{\bibinfo}[2]{#2}
\ifx\xfnm\relax \def\xfnm[#1]{\unskip,\space#1}\fi
%Type = Article
\bibitem[{Antoniadis et~al.(1998)Antoniadis, Arkani-Hamed, Dimopoulos, and
  Dvali}]{Antoniadis1998}
\bibinfo{author}{I.~Antoniadis}, \bibinfo{author}{N.~Arkani-Hamed},
  \bibinfo{author}{S.~Dimopoulos}, \bibinfo{author}{G.~Dvali},
\newblock \bibinfo{title}{New dimensions at a millimeter to a fermi and
  superstrings at a tev},
\newblock \bibinfo{journal}{Phys. Lett.} \bibinfo{volume}{B 436}
  (\bibinfo{year}{1998}) \bibinfo{pages}{257}.
%Type = Article
\bibitem[{Randall and Sundrum(1999{\natexlab{a}})}]{Randall1999a}
\bibinfo{author}{L.~Randall}, \bibinfo{author}{R.~Sundrum},
\newblock \bibinfo{title}{A large mass hierarchy from a small extra dimension},
\newblock \bibinfo{journal}{Phys. Rev. Lett.} \bibinfo{volume}{83}
  (\bibinfo{year}{1999}{\natexlab{a}}) \bibinfo{pages}{3370}.
%Type = Article
\bibitem[{Randall and Sundrum(1999{\natexlab{b}})}]{Randall1999b}
\bibinfo{author}{L.~Randall}, \bibinfo{author}{R.~Sundrum},
\newblock \bibinfo{title}{An alternative to compactification},
\newblock \bibinfo{journal}{Phys. Rev. Lett.} \bibinfo{volume}{83}
  (\bibinfo{year}{1999}{\natexlab{b}}) \bibinfo{pages}{4690}.
%Type = Article
\bibitem[{Rubakov and Shaposhnikov(1983)}]{Rubakov2}
\bibinfo{author}{V.~Rubakov}, \bibinfo{author}{M.~Shaposhnikov},
\newblock \bibinfo{title}{Extra space-time dimensions: towards a solution to
  the cosmological constant problem},
\newblock \bibinfo{journal}{Phys. Lett} \bibinfo{volume}{B 125}
  (\bibinfo{year}{1983}) \bibinfo{pages}{139}.
%Type = Article
\bibitem[{Arkani-Hamed et~al.(1998)Arkani-Hamed, Dimopoulos, and
  Dvali}]{ArkaniHamed1998rs}
\bibinfo{author}{N.~Arkani-Hamed}, \bibinfo{author}{S.~Dimopoulos},
  \bibinfo{author}{G.~Dvali},
\newblock \bibinfo{title}{The hierarchy problem and new dimensions at a
  millimeter},
\newblock \bibinfo{journal}{Phys. Lett.} \bibinfo{volume}{B 429}
  (\bibinfo{year}{1998}) \bibinfo{pages}{263}.
%Type = Article
\bibitem[{Arkani-Hamed et~al.(2000)Arkani-Hamed, Dimopoulos, Kaloper, and
  Sundrum}]{Cosmological2000}
\bibinfo{author}{N.~Arkani-Hamed}, \bibinfo{author}{S.~Dimopoulos},
  \bibinfo{author}{N.~Kaloper}, \bibinfo{author}{R.~Sundrum},
\newblock \bibinfo{title}{A small cosmological constant from a large extra
  dimension},
\newblock \bibinfo{journal}{Phys. Lett.} \bibinfo{volume}{B 480}
  (\bibinfo{year}{2000}) \bibinfo{pages}{193}.
%Type = Article
\bibitem[{Kim et~al.(2001)Kim, Kyae, and Lee}]{PhysRevLett.86.4223}
\bibinfo{author}{J.~E. Kim}, \bibinfo{author}{B.~Kyae}, \bibinfo{author}{H.~M.
  Lee},
\newblock \bibinfo{title}{Randall-sundrum model for self-tuning the
  cosmological constant},
\newblock \bibinfo{journal}{Phys. Rev. Lett.} \bibinfo{volume}{86}
  (\bibinfo{year}{2001}) \bibinfo{pages}{4223}.
%Type = Article
\bibitem[{Dey et~al.(2009)Dey, Mukhopadhyaya, and SenGupta}]{coscon2009}
\bibinfo{author}{P.~Dey}, \bibinfo{author}{B.~Mukhopadhyaya},
  \bibinfo{author}{S.~SenGupta},
\newblock \bibinfo{title}{Neutrino masses, the cosmological constant and a
  stable universe in a randall-sundrum scenario},
\newblock \bibinfo{journal}{Phys. Rev.} \bibinfo{volume}{D 80}
  (\bibinfo{year}{2009}) \bibinfo{pages}{055029}.
%Type = Article
\bibitem[{Neupane(2011)}]{Neupane2010ey}
\bibinfo{author}{I.~P. Neupane},
\newblock \bibinfo{title}{De sitter brane-world, localization of gravity, and
  the cosmological constant},
\newblock \bibinfo{journal}{Phys. Rev.} \bibinfo{volume}{D 83}
  (\bibinfo{year}{2011}) \bibinfo{pages}{086004}.
%Type = Article
\bibitem[{Haghani et~al.(2012)Haghani, Sepangi, and Shahidi}]{BraneFR2012}
\bibinfo{author}{Z.~Haghani}, \bibinfo{author}{H.~R. Sepangi},
  \bibinfo{author}{S.~Shahidi},
\newblock \bibinfo{title}{Cosmological dynamics of brane f(r) gravity},
\newblock \bibinfo{journal}{JCAP} \bibinfo{volume}{1202} (\bibinfo{year}{2012})
  \bibinfo{pages}{031}.
%Type = Article
\bibitem[{George et~al.(2009)George, Trodden, and Volkas}]{ThickBrane20093}
\bibinfo{author}{D.~P. George}, \bibinfo{author}{M.~Trodden},
  \bibinfo{author}{R.~R. Volkas},
\newblock \bibinfo{title}{Extra-dimensional cosmology with domain-wall branes},
\newblock \bibinfo{journal}{JHEP} \bibinfo{volume}{0902} (\bibinfo{year}{2009})
  \bibinfo{pages}{035}.
%Type = Article
\bibitem[{O.~DeWolfe and Karch(2000)}]{ThickBraneDewolfe}
\bibinfo{author}{S.~G. O.~DeWolfe, D.Z.~Freedman}, \bibinfo{author}{A.~Karch},
\newblock \bibinfo{title}{Modeling the fifth dimension with scalars and
  gravity},
\newblock \bibinfo{journal}{Phys. Rev.} \bibinfo{volume}{D 62}
  (\bibinfo{year}{2000}) \bibinfo{pages}{046008}.
%Type = Article
\bibitem[{Gregory et~al.(2000)Gregory, Rubakov, and
  Sibiryakov}]{Gregory:2000jc}
\bibinfo{author}{R.~Gregory}, \bibinfo{author}{V.~Rubakov},
  \bibinfo{author}{S.~M. Sibiryakov},
\newblock \bibinfo{title}{Opening up extra dimensions at ultra large scales},
\newblock \bibinfo{journal}{Phys. Rev. Lett.} \bibinfo{volume}{84}
  (\bibinfo{year}{2000}) \bibinfo{pages}{5928}.
%Type = Article
\bibitem[{Kaloper et~al.(2000)Kaloper, March-Russell, Starkman, and
  Trodden}]{Kaloper2000jb}
\bibinfo{author}{N.~Kaloper}, \bibinfo{author}{J.~March-Russell},
  \bibinfo{author}{G.~D. Starkman}, \bibinfo{author}{M.~Trodden},
\newblock \bibinfo{title}{Compact hyperbolic extra dimensions: Branes,
  kaluza-klein modes and cosmology},
\newblock \bibinfo{journal}{Phys. Rev. Lett.} \bibinfo{volume}{85}
  (\bibinfo{year}{2000}) \bibinfo{pages}{928}.
%Type = Article
\bibitem[{Wang(2002)}]{PhysRevD.66.024024}
\bibinfo{author}{A.~Wang},
\newblock \bibinfo{title}{Thick de sitter 3-branes, dynamic black holes, and
  localization of gravity},
\newblock \bibinfo{journal}{Phys. Rev.} \bibinfo{volume}{D 66}
  (\bibinfo{year}{2002}) \bibinfo{pages}{024024}.
%Type = Article
\bibitem[{Kobayashi et~al.(2002)Kobayashi, Koyama, and Soda}]{ThickBrane2002}
\bibinfo{author}{S.~Kobayashi}, \bibinfo{author}{K.~Koyama},
  \bibinfo{author}{J.~Soda},
\newblock \bibinfo{title}{Thick brane worlds and their stability},
\newblock \bibinfo{journal}{Phys. Rev.} \bibinfo{volume}{D 65}
  (\bibinfo{year}{2002}) \bibinfo{pages}{064014}.
%Type = Article
\bibitem[{Melfo et~al.(2003)Melfo, Pantoja, and Skirzewski}]{ThickBrane2003}
\bibinfo{author}{A.~Melfo}, \bibinfo{author}{N.~Pantoja},
  \bibinfo{author}{A.~Skirzewski},
\newblock \bibinfo{title}{Thick domain wall space-times with and without
  reflection symmetry},
\newblock \bibinfo{journal}{Phys. Rev.} \bibinfo{volume}{D 67}
  (\bibinfo{year}{2003}) \bibinfo{pages}{105003}.
%Type = Article
\bibitem[{Bazeia and Gomes(2004)}]{Blochbrane2004}
\bibinfo{author}{D.~Bazeia}, \bibinfo{author}{A.~Gomes},
\newblock \bibinfo{title}{Bloch brane},
\newblock \bibinfo{journal}{JHEP} \bibinfo{volume}{0405} (\bibinfo{year}{2004})
  \bibinfo{pages}{012}.
%Type = Article
\bibitem[{Bazeia et~al.(2006)Bazeia, Brito, and Losano}]{ThickBraneBazeia2006}
\bibinfo{author}{D.~Bazeia}, \bibinfo{author}{F.~Brito},
  \bibinfo{author}{L.~Losano},
\newblock \bibinfo{title}{Scalar fields, bent branes, and rg flow},
\newblock \bibinfo{journal}{JHEP} \bibinfo{volume}{0611} (\bibinfo{year}{2006})
  \bibinfo{pages}{064}.
%Type = Article
\bibitem[{Cardoso et~al.(2007)Cardoso, Koyama, Mennim, Seahra, and
  Wands}]{ThickBrane20071}
\bibinfo{author}{A.~Cardoso}, \bibinfo{author}{K.~Koyama},
  \bibinfo{author}{A.~Mennim}, \bibinfo{author}{S.~S. Seahra},
  \bibinfo{author}{D.~Wands},
\newblock \bibinfo{title}{Coupled bulk and brane fields about a de sitter
  brane},
\newblock \bibinfo{journal}{Phys. Rev.} \bibinfo{volume}{D 75}
  (\bibinfo{year}{2007}) \bibinfo{pages}{084002}.
%Type = Article
\bibitem[{V.~Dzhunushaliev and Aguilar-Rudametkin(2008)}]{ThickBrane20082}
\bibinfo{author}{D.~S. V.~Dzhunushaliev, V.~Folomeev},
  \bibinfo{author}{S.~Aguilar-Rudametkin},
\newblock \bibinfo{title}{6d thick branes from interacting scalar fields},
\newblock \bibinfo{journal}{Phys. Rev.} \bibinfo{volume}{D77}
  (\bibinfo{year}{2008}) \bibinfo{pages}{044006}.
%Type = Article
\bibitem[{Liu et~al.(2010)Liu, Yang, and Zhong}]{KeYang2009Weyl}
\bibinfo{author}{Y.-X. Liu}, \bibinfo{author}{K.~Yang},
  \bibinfo{author}{Y.~Zhong},
\newblock \bibinfo{title}{de sitter thick brane solution in weyl geometry},
\newblock \bibinfo{journal}{JHEP} \bibinfo{volume}{1010} (\bibinfo{year}{2010})
  \bibinfo{pages}{069}.
%Type = Article
\bibitem[{Liu et~al.(2011)Liu, Zhong, Zhao, and
  Li}]{ThickBraneZhongYuan2011JHEP}
\bibinfo{author}{Y.-X. Liu}, \bibinfo{author}{Y.~Zhong}, \bibinfo{author}{Z.-H.
  Zhao}, \bibinfo{author}{H.-T. Li},
\newblock \bibinfo{title}{Domain wall brane in squared curvature gravity},
\newblock \bibinfo{journal}{JHEP} \bibinfo{volume}{1106} (\bibinfo{year}{2011})
  \bibinfo{pages}{135}.
%Type = Article
\bibitem[{Liu et~al.(2012)Liu, Yang, Guo, and Zhong}]{Liu:2012rc}
\bibinfo{author}{Y.-X. Liu}, \bibinfo{author}{K.~Yang},
  \bibinfo{author}{H.~Guo}, \bibinfo{author}{Y.~Zhong},
\newblock \bibinfo{title}{Domain wall brane in eddington inspired born-infeld
  gravity},
\newblock \bibinfo{journal}{Phys.Rev.} \bibinfo{volume}{D85}
  (\bibinfo{year}{2012}) \bibinfo{pages}{124053}.
%Type = Article
\bibitem[{Liu et~al.(2013)Liu, Wang, Wu, and Zhong}]{Liu:2012mia}
\bibinfo{author}{Y.-X. Liu}, \bibinfo{author}{Y.-Q. Wang},
  \bibinfo{author}{S.-F. Wu}, \bibinfo{author}{Y.~Zhong},
\newblock \bibinfo{title}{Analytic solutions of brane in critical gravity},
\newblock \bibinfo{journal}{Phys.Rev.} \bibinfo{volume}{D88}
  (\bibinfo{year}{2013}) \bibinfo{pages}{104033}.
%Type = Article
\bibitem[{Bazeia et~al.(2014)Bazeia, Lobao, Menezes, Petrov, and
  da~Silva}]{Bazeia:2013uva}
\bibinfo{author}{D.~Bazeia}, \bibinfo{author}{A.~J. Lobao, A.o},
  \bibinfo{author}{R.~Menezes}, \bibinfo{author}{A.~Y. Petrov},
  \bibinfo{author}{A.~da~Silva},
\newblock \bibinfo{title}{Braneworld solutions for f(r) models with
  non-constant curvature},
\newblock \bibinfo{journal}{Phys. Lett.} \bibinfo{volume}{B 729}
  (\bibinfo{year}{2014}) \bibinfo{pages}{127}.
%Type = Article
\bibitem[{Gherghetta et~al.(2000)Gherghetta, Roessl, and
  Shaposhnikov}]{Gherghetta:2000jf}
\bibinfo{author}{T.~Gherghetta}, \bibinfo{author}{E.~Roessl},
  \bibinfo{author}{M.~E. Shaposhnikov},
\newblock \bibinfo{title}{Living inside a hedgehog: Higher dimensional
  solutions that localize gravity},
\newblock \bibinfo{journal}{Phys. Lett.} \bibinfo{volume}{B 491}
  (\bibinfo{year}{2000}) \bibinfo{pages}{353}.
%Type = Article
\bibitem[{Zhong et~al.(2011)Zhong, Liu, and Yang}]{Zhong:2010ae}
\bibinfo{author}{Y.~Zhong}, \bibinfo{author}{Y.-X. Liu},
  \bibinfo{author}{K.~Yang},
\newblock \bibinfo{title}{Tensor perturbations of $f(r)$-branes},
\newblock \bibinfo{journal}{Phys. Lett.} \bibinfo{volume}{B 699}
  (\bibinfo{year}{2011}) \bibinfo{pages}{398}.
%Type = Article
\bibitem[{Witten(1998)}]{Witten:1998qj}
\bibinfo{author}{E.~Witten},
\newblock \bibinfo{title}{Anti-de sitter space and holography},
\newblock \bibinfo{journal}{Adv. Theor. Math. Phys.} \bibinfo{volume}{2}
  (\bibinfo{year}{1998}) \bibinfo{pages}{253}.
%Type = Article
\bibitem[{Gubser et~al.(1998)Gubser, Klebanov, and Polyakov}]{Gubser:1998bc}
\bibinfo{author}{S.~Gubser}, \bibinfo{author}{I.~R. Klebanov},
  \bibinfo{author}{A.~M. Polyakov},
\newblock \bibinfo{title}{Gauge theory correlators from noncritical string
  theory},
\newblock \bibinfo{journal}{Phys. Lett.} \bibinfo{volume}{B 428}
  (\bibinfo{year}{1998}) \bibinfo{pages}{105}.
%Type = Article
\bibitem[{Maldacena(1998)}]{Maldacena:1997re}
\bibinfo{author}{J.~M. Maldacena},
\newblock \bibinfo{title}{The large n limit of superconformal field theories
  and supergravity},
\newblock \bibinfo{journal}{Adv. Theor. Math. Phys.} \bibinfo{volume}{2}
  (\bibinfo{year}{1998}) \bibinfo{pages}{231}.
%Type = Article
\bibitem[{Grossman and Neubert(2000)}]{Grossman:1999ra}
\bibinfo{author}{Y.~Grossman}, \bibinfo{author}{M.~Neubert},
\newblock \bibinfo{title}{Neutrino masses and mixings in nonfactorizable
  geometry},
\newblock \bibinfo{journal}{Phys. Lett.} \bibinfo{volume}{B 474}
  (\bibinfo{year}{2000}) \bibinfo{pages}{361}.
%Type = Article
\bibitem[{Bajc and Gabadadze(2000)}]{Bajc:1999mh}
\bibinfo{author}{B.~Bajc}, \bibinfo{author}{G.~Gabadadze},
\newblock \bibinfo{title}{Localization of matter and cosmological constant on a
  brane in anti-de sitter space},
\newblock \bibinfo{journal}{Phys. Lett.} \bibinfo{volume}{B 474}
  (\bibinfo{year}{2000}) \bibinfo{pages}{282}.
%Type = Article
\bibitem[{Gremm(2000)}]{Gremm:1999pj}
\bibinfo{author}{M.~Gremm},
\newblock \bibinfo{title}{Four-dimensional gravity on a thick domain wall},
\newblock \bibinfo{journal}{Phys. Lett.} \bibinfo{volume}{B 478}
  (\bibinfo{year}{2000}) \bibinfo{pages}{434}.
%Type = Article
\bibitem[{Chang et~al.(2000)Chang, Hisano, Nakano, Okada, and
  Yamaguchi}]{Chang1999nh}
\bibinfo{author}{S.~Chang}, \bibinfo{author}{J.~Hisano},
  \bibinfo{author}{H.~Nakano}, \bibinfo{author}{N.~Okada},
  \bibinfo{author}{M.~Yamaguchi},
\newblock \bibinfo{title}{Bulk standard model in the randall-sundrum
  background},
\newblock \bibinfo{journal}{Phys. Rev.} \bibinfo{volume}{D 62}
  (\bibinfo{year}{2000}) \bibinfo{pages}{084025}.
%Type = Article
\bibitem[{Randjbar-Daemi and Shaposhnikov(2000)}]{RandjbarDaemi:2000cr}
\bibinfo{author}{S.~Randjbar-Daemi}, \bibinfo{author}{M.~E. Shaposhnikov},
\newblock \bibinfo{title}{Fermion zero modes on brane worlds},
\newblock \bibinfo{journal}{Phys. Lett.} \bibinfo{volume}{B 492}
  (\bibinfo{year}{2000}) \bibinfo{pages}{361}.
%Type = Article
\bibitem[{Kehagias and Tamvakis(2001)}]{Kehagias:2000au}
\bibinfo{author}{A.~Kehagias}, \bibinfo{author}{K.~Tamvakis},
\newblock \bibinfo{title}{Localized gravitons, gauge bosons and chiral fermions
  in smooth spaces generated by a bounce},
\newblock \bibinfo{journal}{Phys. Lett.} \bibinfo{volume}{B 504}
  (\bibinfo{year}{2001}) \bibinfo{pages}{38}.
%Type = Article
\bibitem[{Oda(2001)}]{Oda2001}
\bibinfo{author}{I.~Oda},
\newblock \bibinfo{title}{Localization of bulk fields on ads(4) brane in
  ads(5)},
\newblock \bibinfo{journal}{Phys. Lett.} \bibinfo{volume}{B 508}
  (\bibinfo{year}{2001}) \bibinfo{pages}{96}.
%Type = Article
\bibitem[{Ringeval et~al.(2002)Ringeval, Peter, and Uzan}]{Ringeval:2001cq}
\bibinfo{author}{C.~Ringeval}, \bibinfo{author}{P.~Peter},
  \bibinfo{author}{J.-P. Uzan},
\newblock \bibinfo{title}{Localization of massive fermions on the brane},
\newblock \bibinfo{journal}{Phys. Rev.} \bibinfo{volume}{D 65}
  (\bibinfo{year}{2002}) \bibinfo{pages}{044016}.
%Type = Article
\bibitem[{Ichinose(2002)}]{Ichinose:2002kg}
\bibinfo{author}{S.~Ichinose},
\newblock \bibinfo{title}{Fermions in kaluza-klein and randall-sundrum
  theories},
\newblock \bibinfo{journal}{Phys. Rev.} \bibinfo{volume}{D 66}
  (\bibinfo{year}{2002}) \bibinfo{pages}{104015}.
%Type = Article
\bibitem[{Koley and Kar(2005)}]{Koley:2004at}
\bibinfo{author}{R.~Koley}, \bibinfo{author}{S.~Kar},
\newblock \bibinfo{title}{Scalar kinks and fermion localisation in warped
  spacetimes},
\newblock \bibinfo{journal}{Class. Quant. Grav.} \bibinfo{volume}{22}
  (\bibinfo{year}{2005}) \bibinfo{pages}{753}.
%Type = Article
\bibitem[{Davies et~al.(2008)Davies, George, and Volkas}]{Davies:2007xr}
\bibinfo{author}{R.~Davies}, \bibinfo{author}{D.~P. George},
  \bibinfo{author}{R.~R. Volkas},
\newblock \bibinfo{title}{The standard model on a domain-wall brane},
\newblock \bibinfo{journal}{Phys. Rev.} \bibinfo{volume}{D 77}
  (\bibinfo{year}{2008}) \bibinfo{pages}{124038}.
%Type = Article
\bibitem[{Liu et~al.(2008)Liu, Zhang, Wei, and Duan}]{Liu2008WeylPT}
\bibinfo{author}{Y.-X. Liu}, \bibinfo{author}{L.-D. Zhang},
  \bibinfo{author}{S.-W. Wei}, \bibinfo{author}{Y.-S. Duan},
\newblock \bibinfo{title}{Localization and mass spectrum of matters on weyl
  thick branes},
\newblock \bibinfo{journal}{JHEP} \bibinfo{volume}{0808} (\bibinfo{year}{2008})
  \bibinfo{pages}{041}.
%Type = Article
\bibitem[{Zhao et~al.(2010)Zhao, Liu, Li, and Wang}]{Zhao:2010mk}
\bibinfo{author}{Z.-H. Zhao}, \bibinfo{author}{Y.-X. Liu},
  \bibinfo{author}{H.-T. Li}, \bibinfo{author}{Y.-Q. Wang},
\newblock \bibinfo{title}{Effects of the variation of mass on fermion
  localization and resonances on thick branes},
\newblock \bibinfo{journal}{Phys. Rev.} \bibinfo{volume}{D 82}
  (\bibinfo{year}{2010}) \bibinfo{pages}{084030}.
%Type = Article
\bibitem[{Jones et~al.(2013)Jones, Munoz, Singleton, and
  Triyanta}]{Jones:2013ofa}
\bibinfo{author}{P.~Jones}, \bibinfo{author}{G.~Munoz},
  \bibinfo{author}{D.~Singleton}, \bibinfo{author}{Triyanta},
\newblock \bibinfo{title}{Field localization and nambu jona-lasinio mass
  generation mechanism in an alternative 5-dimensional brane model},
\newblock \bibinfo{journal}{Phys. Rev.} \bibinfo{volume}{D 88}
  (\bibinfo{year}{2013}) \bibinfo{pages}{025048}.
%Type = Article
\bibitem[{Xie et~al.(2013)Xie, Yang, and Zhao}]{Xie2013rka}
\bibinfo{author}{Q.-Y. Xie}, \bibinfo{author}{J.~Yang},
  \bibinfo{author}{L.~Zhao},
\newblock \bibinfo{title}{Resonance mass spectra of gravity and fermion on
  bloch branes},
\newblock \bibinfo{journal}{Phys. Rev.} \bibinfo{volume}{D 88}
  (\bibinfo{year}{2013}) \bibinfo{pages}{105014}.
%Type = Article
\bibitem[{Cembranos et~al.(2013)Cembranos, Delgado, and
  Dobado}]{Cembranos:2013qja}
\bibinfo{author}{J.~A.~R. Cembranos}, \bibinfo{author}{R.~L. Delgado},
  \bibinfo{author}{A.~Dobado},
\newblock \bibinfo{title}{Brane-worlds at the lhc: Branons and kk-gravitons},
\newblock \bibinfo{journal}{Phys. Rev.} \bibinfo{volume}{D 88}
  (\bibinfo{year}{2013}) \bibinfo{pages}{075021}.
%Type = Article
\bibitem[{Costa et~al.(2013)Costa, Silva, and Almeida}]{Costa:2013eua}
\bibinfo{author}{F.~Costa}, \bibinfo{author}{J.~Silva},
  \bibinfo{author}{C.~Almeida},
\newblock \bibinfo{title}{Gauge vector field localization on 3-brane placed in
  a warped transverse resolved conifold},
\newblock \bibinfo{journal}{Phys. Rev.} \bibinfo{volume}{D 87}
  (\bibinfo{year}{2013}) \bibinfo{pages}{125010}.
%Type = Article
\bibitem[{Duff and Liu(2001)}]{Duff2000se}
\bibinfo{author}{M.~Duff}, \bibinfo{author}{J.~T. Liu},
\newblock \bibinfo{title}{Hodge duality on the brane},
\newblock \bibinfo{journal}{Phys. Lett.} \bibinfo{volume}{B 508}
  (\bibinfo{year}{2001}) \bibinfo{pages}{381}.
%Type = Article
\bibitem[{Ma and Yeh(2013)}]{Ma:2012hc}
\bibinfo{author}{C.-T. Ma}, \bibinfo{author}{C.-H. Yeh},
\newblock \bibinfo{title}{Supersymmetry and bps states on d4-brane in large
  c-field background},
\newblock \bibinfo{journal}{JHEP} \bibinfo{volume}{1303} (\bibinfo{year}{2013})
  \bibinfo{pages}{131}.
%Type = Article
\bibitem[{Ho and Ma(2013)}]{Ho:2013paa}
\bibinfo{author}{P.-M. Ho}, \bibinfo{author}{C.-T. Ma},
\newblock \bibinfo{title}{Effective action for dp-brane in large rr (p-1)-form
  background},
\newblock \bibinfo{journal}{JHEP} \bibinfo{volume}{1305} (\bibinfo{year}{2013})
  \bibinfo{pages}{056}.
%Type = Article
\bibitem[{Mukhopadhyaya et~al.(2002)Mukhopadhyaya, Sen, and
  SenGupta}]{PRLKR2002}
\bibinfo{author}{B.~Mukhopadhyaya}, \bibinfo{author}{S.~Sen},
  \bibinfo{author}{S.~SenGupta},
\newblock \bibinfo{title}{Does a randall-sundrum scenario create the illusion
  of a torsion free universe?},
\newblock \bibinfo{journal}{Phys. Rev. Lett.} \bibinfo{volume}{89}
  (\bibinfo{year}{2002}) \bibinfo{pages}{121101}.
%Type = Article
\bibitem[{Mukhopadhyaya et~al.(2004)Mukhopadhyaya, Sen, Sen, and
  SenGupta}]{Mukhopadhyaya2004}
\bibinfo{author}{B.~Mukhopadhyaya}, \bibinfo{author}{S.~Sen},
  \bibinfo{author}{S.~Sen}, \bibinfo{author}{S.~SenGupta},
\newblock \bibinfo{title}{Bulk kalb-ramond field in randall-sundrum scenario},
\newblock \bibinfo{journal}{Phys. Rev.} \bibinfo{volume}{D 70}
  (\bibinfo{year}{2004}) \bibinfo{pages}{066009}.
%Type = Article
\bibitem[{Mukhopadhyaya et~al.(2007)Mukhopadhyaya, Sen, and SenGupta}]{QformRS}
\bibinfo{author}{B.~Mukhopadhyaya}, \bibinfo{author}{S.~Sen},
  \bibinfo{author}{S.~SenGupta},
\newblock \bibinfo{title}{Bulk antisymmetric tensor fields in a randall-sundrum
  model},
\newblock \bibinfo{journal}{Phys. Rev.} \bibinfo{volume}{D 76}
  (\bibinfo{year}{2007}) \bibinfo{pages}{121501}.
%Type = Article
\bibitem[{Alencar et~al.(2010{\natexlab{a}})Alencar, Tahim, Landim, Muniz, and
  Costa~Filho}]{qRSdilation-eprintv2}
\bibinfo{author}{G.~Alencar}, \bibinfo{author}{M.~Tahim},
  \bibinfo{author}{R.~Landim}, \bibinfo{author}{C.~Muniz},
  \bibinfo{author}{R.~Costa~Filho},
\newblock \bibinfo{title}{Bulk antisymmetric tensor fields coupled to a dilaton
  in a randall-sundrum model},
\newblock \bibinfo{journal}{Phys. Rev.} \bibinfo{volume}{D 82}
  (\bibinfo{year}{2010}{\natexlab{a}}) \bibinfo{pages}{104053}.
%Type = Article
\bibitem[{Alencar et~al.(2010{\natexlab{b}})Alencar, Landim, Tahim, Muniz, and
  Costa~Filho}]{QformThickRS}
\bibinfo{author}{G.~Alencar}, \bibinfo{author}{R.~Landim},
  \bibinfo{author}{M.~Tahim}, \bibinfo{author}{C.~Muniz},
  \bibinfo{author}{R.~Costa~Filho},
\newblock \bibinfo{title}{Antisymmetric tensor fields in randall sundrum thick
  branes},
\newblock \bibinfo{journal}{Phys. Lett.} \bibinfo{volume}{B 693}
  (\bibinfo{year}{2010}{\natexlab{b}}) \bibinfo{pages}{503}.
%Type = Article
\bibitem[{Fu et~al.(2012)Fu, Liu, Yang, and Wei}]{Fu2012sa}
\bibinfo{author}{C.-E. Fu}, \bibinfo{author}{Y.-X. Liu},
  \bibinfo{author}{K.~Yang}, \bibinfo{author}{S.-W. Wei},
\newblock \bibinfo{title}{q-form fields on p-branes},
\newblock \bibinfo{journal}{JHEP} \bibinfo{volume}{1210} (\bibinfo{year}{2012})
  \bibinfo{pages}{060}.
%Type = Article
\bibitem[{Koivisto and Nunes(2013)}]{PhysRevD.88.123512}
\bibinfo{author}{T.~S. Koivisto}, \bibinfo{author}{N.~J. Nunes},
\newblock \bibinfo{title}{Coupled three-form dark energy},
\newblock \bibinfo{journal}{Phys. Rev. D} \bibinfo{volume}{88}
  (\bibinfo{year}{2013}) \bibinfo{pages}{123512}.
%Type = Book
\bibitem[{Mannheim(2005)}]{Mannheim2005}
\bibinfo{author}{P.~D. Mannheim}, \bibinfo{title}{Brane-Localized Gravity},
  \bibinfo{publisher}{World Scientific Publishing Company},
  \bibinfo{year}{2005}.

\end{thebibliography}

\end{document}